\documentclass[aps,a4paper,showpacs,prd,floatfix,twocolumn,amsmath,amssymb,nofootinbib]{revtex4-1}
\usepackage{graphicx}
\usepackage{epsfig}
\usepackage[normalem]{ulem}
\usepackage{bm}
\usepackage{soul} 
\usepackage{multirow} 
\usepackage{color, colortbl}
\usepackage{numprint} 

\def\l{\left(} 
\def\r{\right)} 
\def\d3{\mathrm{d}^3}
\def\d4{\mathrm{d}^4}
\def\rd{\mathrm{d}} 
 
\def\be{\begin{equation}}
\def\ee{\end{equation}}

\begin{document}

\title{Neutron star inner crust: effects of rotation and magnetic fields }

\author{Ivo Sengo$^{1}$}
\email{ivoabs@gmail.com}
\author{Helena Pais$^{1}$}
\email{{ hpais@uc.pt}}
\author{Bruno Franzon $^{2}$}
\email{franzon@fias.uni-frankfurt.de}
\author{Constan\c ca Provid\^encia$^{1}$}
\email{cp@uc.pt}

\affiliation{$^{1}$CFisUC, Department of Physics, University of Coimbra, 3004-516 Coimbra, Portugal.\\
$^{2}$Frankfurt Institute for Advanced Studies, Ruth-Moufang-1, 60438 Frankfurt am Main, Germany.}

\begin{abstract}

We study the role of the pasta phases on the properties of rotating and magnetized neutron stars. In order to investigate such systems, we make use of two different relativistic mean-field unified inner-crust--core equations of state, with a different density dependence of the symmetry energy, and an inner-crust computed within a Thomas-Fermi calculation. Special attention is given to the crust-core transition density, and the pasta phases effects on the global properties of stars. The effects of strong magnetic fields and fast rotation are computed by solving the Einstein-Maxwell equations self-consistently, taking into account anisotropies induced by the centrifugal and the Lorentz force. The location of the magnetic field neutral line and the  maximum of the Lorentz force on the equatorial plane are calculated. The conditions under which they fall inside the inner crust region are discussed. We verified that models with a larger symmetry energy  slope show more  sensitivity to the variation of the magnetic field. One of the maxima of the Lorentz force, as well as the neutral line, and for a certain range of frequencies, fall inside the inner crust region. This may have consequences in the fracture of the crust, and may help explain phenomena associated with star quakes.

\end{abstract}

\maketitle

\section{Introduction}

Neutron stars (NS) are not only extremely dense objects,
but they are known to be associated with strong magnetic fields, and fast rotation as well. At present, it is commonly accepted that the huge range of densities inside NS can be naturally divided into several regions. Typically, the neutron star structure can be divided into an outer crust, an inner crust and a core.  
The outer crust region of neutron stars has an equation of state relatively well-known \cite{bps,HP,Ruester2005}. The same is not true for the inner crust. 
This region of the star begins when neutrons start  dripping out of the nuclei at densities of about $\rho_{{\rm drip}}\sim 4.3\times 10^{11}$ g/cm$^{3}$.  As a result, the inner crust is formed by very neutron-rich nuclei, immersed in a gas of neutrons and electrons. Heavy clusters, the pasta phases, form due to the competition between the nuclear and Coulomb forces \cite{Ravenhall1983,Pais2012PRL,Pais2016vlasov, watanabe2005simulation}.  This may affect the cooling of the proton neutron star.

Pulsars rotate extremely fast, which is related to their formation \cite{Lorimer2008}. As the star core collapses, its rotation rate increases as a result of conservation of angular momentum, hence, pulsars rotate up to several hundred times per second. In the case of millisecond pulsars, they are thought to achieve  such high speeds because they are gravitationally bound in a binary system with another star. During part of their life, matter flows from the companion star to the pulsar. Over time, the impact of the accreted matter spins up the pulsar's rotation. 

In addition, classes of neutron stars known as magnetars have strong surface magnetic fields that span the range $\sim 10^{12-15}$ G. Such fields are usually estimated from observations of the star's period, and  period derivative. One expects to find even stronger magnetic fields inside these stars. According to  the virial theorem, which gives an upper estimate for the magnetic field inside neutron stars, they can possess stronger central magnetic fields, of the order of  $\sim 10^{18}$ G  \cite{lai1991cold, cardall2001effects}.

The main objective of the present work is to understand how the
distribution of the poloidal magnetic field lines affect the
inner crust of a neutron star.  Moreover, we want precisely to identify the
thickness of the crust and the position  of  the poloidal neutral line
with  respect to the crust, taking as reference  an unified equation of
state, and allowing for the symmetry  energy to vary. The knowledge of
the size and position of the crust is important to understand its
possible role in the stabilization of the magnetic field and the low
frequency  quasi-periodic oscillations (QPO) associated with  magnetar
flares \cite{Steiner2009,Gabler2012,Sotani2013,Deibel2013,Sotani2019}.

It has been suggested that  QPO
observed in the decay tails of magnetar flares result  from  seismic
vibrations from neutron stars. Some of these oscillations may be
confined to the crust, in particular the low frequency  ones,  and,  in this case, they are perfect  probes of
the crust  EoS, as discussed in \cite{Steiner2009}. The frequency of these modes
 is directly related  with  both the thickness of the crust, and  the  density-dependence of the symmetry energy \cite{Sotani2013,Deibel2013,Sotani2019}. Another possible interpretation is the association of QPO to magneto-elastic modes \cite{Gabler2012}.

Recently, the  evolution of the magnetic field structure during the
late stage of a proto-neutron star  has been studied
\cite{Lander2020}. It was shown that the structure of the magnetic
field is similar in a hot and cold NS, the poloidal component of the
field being stronger than the toroidal  one. 
Instabilities may originate a large release of the magnetic energy,
but then it is difficult to explain the strong magnetic fields
that many magnetars have.  The authors suggest that one of the
possible mechanisms to stabilize the magnetic field is the
solidification of the crust, starting at the crust-core
transition. The formation of a solid crust would give rise to elastic
forces  that would avoid the development of magnetic field
instabilities, and a fast decay of the magnetic field. In the present
work, using a realistic unified EoS,  we will  show that the neutral
line of the poloidal field, i.e. the  region where instabilities
develop, may, in fact, fall  in the crust region for a rotating
star. As it will be discussed, this result is sensitive to the density-dependence  of the  symmetry   energy.

We will, therefore, concentrate our attention on
the inner crust-core transition, and will not investigate the outer
crust and transition to the inner crust of a strongly magnetized
star. Several studies have already shown the important effects of the
magnetic field on the outer crust and neutron drip line \cite{Potekhin2013,Chamel2012,Chamel2015,Stein2016}.
In Ref.~\cite{Chatterjee2014}, the authors have shown that including magnetic field effects in the EoS did not affect much the magnetized neutron star structure, therefore, in the following, we consider a non-magnetized EoS.

In order to describe the neutron star interior, the complete stellar matter EoS will be constructed by taking a standard EoS for the outer crust \cite{bps}, with  an adequate inner crust EoS that matches the outer crust EoS at the neutron drip line, and the core EoS at the crust-core transition density \cite{Pais2016vlasov}. Between the neutron drip density and the crust-core transition density, we  employ an inner crust EoS, that we have determined  from a Thomas-Fermi calculation for the NL3 family \cite{Pais2016vlasov},  with the inclusion of the $\omega\rho$ meson coupling terms.  There, the authors addressed the effect of the nonlinear $\omega\rho$ coupling terms on the crust-core transition density and pressure, and on the macroscopic properties of hadronic stars.  We will also consider that the magnetic field affects the extension of the inner crust, as proposed in \cite{Fang2016,Fang2017,Fang2017a,Chen2017}. The  complete EoS will be used as input to determine the star properties, such as the mass and radius, from the integration of the Einstein-Maxwell equations, in order to obtain both rotating and magnetized stellar models \cite{bonazzola1993,Bocquet:1995je,Franzon:2015sya,Franzon:2016urz}.

Pasta phases impact not only the structure of NS, but also may affect their rotation behavior, and the magnetic field distribution. The effects of rotation and strong magnetic fields in the inner crust region, where the pasta phases appear, are going to be analysed, for two model with different slopes of the symmetry energy. 
For this purpose, we are  going to use the Lorene C++ library for numerical relativity\footnote{www.lorene.obspm.fr} to self-consistently study the effects of strong magnetic fields and rotation on neutron stars. We will solve  numerically the coupled Maxwell-Einstein equations by means of a pseudo-spectral method, taking into consideration the anisotropy of the energy-momentum tensor due to the magnetic field, and also the effects of the centrifugal force induced by rotation.

If the NS has a poloidal field with closed lines inside, instabilities will appear in the neighborhood  of the neutral line characterized by a zero magnetic field \cite{Markey1973,Lander2011}, and a mixed poloidal-toroidal configuration will stabilize the NS \cite{Haskell2008,Lander2009,Lander2011,Pili2017,Uryu2019}. However, the relative magnitude of each field component depends on the boundary conditions imposed on the magnetic field  \cite{Haskell2008,Lander2009}, and these will certainly depend on the properties of matter at the NS surface. 
In this paper, we want to address this issue, and the localization of the neutral line of the poloidal magnetic field relative to the crust of the NS will be determined. The structure of the paper is the following: in Sec. \ref{sec2}, we review the formalism, in Sec. \ref{sec4}, the results are presented, and, finally, in Sec. \ref{conclusions} some conclusions are drawn.

\section{Description of Magnetized and Rotating Neutron Stars}
\label{sec2}

In this section, we review the formalism introduced in Refs. \cite{bonazzola1993} and \cite{Bocquet:1995je}, upon which the LORENE code is based. 

Assuming Maximum-Slice Quasi-Isotropic (MSQI) coordinates, stationarity and axisymmetry, the metric tensor reads  

\begin{align}
\begin{split} 
\rd s^2 =  g_{\mu\nu} \rd x^{\mu}x^{\nu}= &-N^{2} \rd t^{2} + A^2( \rd r^2 + r^2 \rd\theta^2) \\
& + B^{2}r^{2}\sin^{2}\theta( \rd \phi - N^{\phi} \rd t)^{2} 
\end{split} \quad , 
\label{METRIC}
\end{align}

\noindent with $N(r,\theta)$, $A(r,\theta)$,  $B(r,\theta)$ and $N^{\phi}(r,\theta)$ function only of $(r,\theta)$.

We only consider stars with poloidal magnetic fields. In this case, the magnetic vector potential $A_{\mu}$ has  components $A_{\mu}=\l A_{t}, 0, 0, A_{\phi} \r$. Note that in Ref.~\cite{frieben2012equilibrium}, the authors constructed toroidal magnetic fields with the choice $A_{\mu}=(0, A_{r}, A_{\theta},0 )$.

One important question about magnetic field in neutron stars is its decay due to dissipation. Hence, stationary models of neutron stars in magnetic fields require a separation of dynamical and dissipative  timescales,  encoded  in  an  assumption  of  infinite conductivity (magnetic fields are 'frozen in' and carried with the  fluid,  a  common  assumption  in  astrophysics).  This  assumption  is  exceedingly  well justified  for neutron star matter, since the ohmic dissipation timescale is  larger than the age of the universe and, therefore, the electric current in the fluid would not suffer ohmic decay  \cite{goldreich1992magnetic}.
Therefore, we assume infinite conductivity inside the stars. In this case, the magnetic flux $ BR^{2}$ ($R$ being the stellar radius) is conserved, and the electric field as measured by the co-moving observer is zero. As a result, we find the relation between the magnetic vector components:
\be
A_{t}=-\Omega A_{\phi} + C \quad ,
\label{AtAphi}
\ee
with $\Omega$ the rotation velocity of the star, and $C$ a constant that determines the total electric charge of the star.

The energy-momentum conservation equation $ \mathsf{\nabla_{\mu} T^{\mu\nu} =0}$ gives an equation of stationary motion for the fluid with magnetic field
\be 
\frac{1}{\mathcal{E}+P} \frac{\partial \,P}{\partial x_{i}}  + \frac{\partial\, \ln \,N}{\partial x_{i}}  - \frac{\ln \,{\rm{\Gamma}}}{\partial x_{i}}  + \frac{F^{i \nu}\,j_{\nu}}{\mathcal{E}+P}=0 \quad ,
\label{EQM} 
\ee
with the spatial coordinates $x_{i}=(r, \theta)$. The first term in Eq. \eqref{EQM} corresponds to the purely matter contribution, the second represents the gravitational potential, the third accounts for the centrifugal effects due to rotation, and the last one is the Lorentz force ($f^{\mu}= F^{\mu\nu}\,j_{\nu}$) induced by  magnetic fields, which, in our case, are generated by the four-electric current $j_{\nu}$. Since $A_{\mu}=(A_{t}, 0,0, A_{\phi})$, then $ j_{\nu}=(j_{t}, 0,0, j_{\phi})$, which comes from the assumption of circularity condition. In other words, there are not meridional currents. 

Eq. \eqref{EQM} is the relativistic version of the Euler equation. One can show, by taking  the rotational, that the Lorentz term in Eq. \eqref{EQM} can be written as
\begin{align}
	\frac{\partial M}{\partial x_{i}} =   \frac{F^{i \nu}\,j_{\nu}}{\mathcal{E}+P}=\left(\frac{ j^{\phi} -\rm{\Omega}\, j^{t}}{\mathcal{E}+P}\right)\frac{\partial A_{\phi}}{\partial x_{i}} \,.
	\label{LORENTZ}
\end{align}  

Note that Eq.\eqref{LORENTZ} represents also the integrability condition of Eq.\eqref{EQM}. The term in parenthesis in Eq. \eqref{LORENTZ} can be a constant, or a function of the magnetic vector potential, $g\left( A_{\phi} \right)$. The arbitrary function $M$ can then be chosen such that:
\be 
\frac{\partial M}{\partial A_{\phi}}= g(A_{\phi}) \quad .
\label{MAG}
\ee

In other words,
\be 
M= M (A_{\phi}(r, \theta)) =\int^{A_{\phi}}_{0} g(u)\,d u  \quad .
\label{MAG1}
\ee
The function $g(u)$ is called the current function, and $M$ is the magnetic potential. Here, the  magnetic  star  models are  obtained  by  assuming a constant value  for the dimensionless current function, also referred to as current function amplitude (CFA), and denoted by $k_{0}$. In Ref. \cite{Bocquet:1995je}, other choices for $g(u)$ were considered, other than constants functions, but the general conclusions remain the same. 

For higher values of the current function, the magnetic field in the star increases proportionally. In addition, $\mathsf{k_{0}}$  is related to the macroscopic electric current via:
\be 
j^{\phi}= {\rm\Omega}\,j^{t} + (\mathcal{E} + P)\,k_{0}  \quad ,
\label{K0}
\ee
which is obtained relating Eq. \eqref{MAG} with Eq. \eqref{LORENTZ}. Here, $\mathcal{E}$ is the energy density and $P$ is the pressure.

Finally, the integral form of the equation of motion for a fluid in the presence of magnetic fields, Eq. \eqref{EQM}, reads:
\be 
H(r, \theta) + \ln \, N(r, \theta) - \ln \,{\rm{\Gamma (r, \theta)}}\, + M(r, \theta) = const. \quad ,
\label{EQMFINAL} 
\ee
where $M$ is the magnetic potential, see Eq.~\eqref{MAG1}, and $H$ is the dimensionless log-enthalpy  (also called pseudo-enthalpy or heat function) defined as 
\be 
H(P) = \int_{0}^{P} \frac{\rd P'}{\mathcal{E}(P')  +  P' } \quad ,
\label{H}
\ee
which can be cast in terms of the specific enthalpy $\mathsf{h}$
\be 
h(P) =\frac{\mathcal{E}(P)  +  P }{m_{b}\,n_{b}} \quad, 
\ee
as
\be 
H(P):= \ln \, h(P) = \ln \, \left( \frac{\mu}{m_{b}}\right) \quad,
\label{HFINAL}
\ee
where $m_{B}=939$ MeV is the baryonic mass, and $\mu$ the baryonic chemical potential.

\section{Results}
\label{sec4}

In the following, we  present the main results of our study. We consider the effect of the magnetic field on the NS crust for a non-rotating star in Sec. \ref{magnetized}, and, for a rotating star,  in  Sec. \ref{rotating}.

\subsection{  Magnetised neutron stars}
\label{magnetized}

\begin{figure}[!hbtp]
	\centering
	\includegraphics[width=0.45\textwidth]{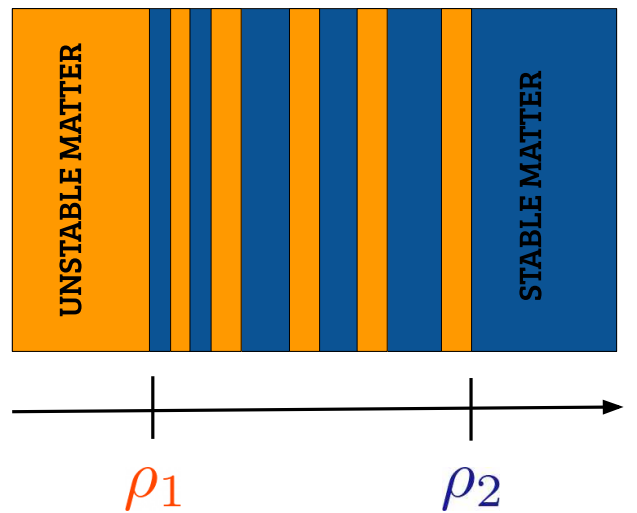}
	\caption{\label{scheme1}(Color online) The extended crust region. The densities $\rho_1$ and $\rho_2$ define the boundaries of this region.}
\end{figure}

As already discussed in Ref. \cite{Fang2017}, the presence of strong magnetic fields originates a region, at the boundary between the inner crust and the core, where homogeneous and non-homogeneous matter (matter with the presence of clusters) coexist -- the extended crust -- identified by the densities $\rho_1$  and $\rho_2$ (cf. Fig. \ref{scheme1}). We shall denote the radii that correspond to each of these densities as $R_1$ and $R_2$, respectively. In this notation, the thickness of the extended crust is defined as $\Delta R_t = R_1 - R_2$, whilst the total size of the crust is given by the difference $\Delta R_2 = R - R_2$ (with $R$ being the coordinate radius of the star). The difference $\Delta R_1 = R-R_1$ corresponds to the size of the crust without the extended region. 

For the region between the surface and  the boundary defined by
   $R_1$ and the density $\rho_1$, which coincides with the crust-core
   transition of a non-magnetized star, we take the  EoS of
   non-magnetized matter.   In \cite{Fang2017a}, it has been shown
   that the magnetic field does not affect much the value of $R_1$, and
   the results of \cite{Lima2013}  concerning the inner crust seem to
   indicate that the structure of the pasta phases inside $R_1$ are
   not influenced by the magnetic field, if the intensity of the field
   satisfies $B<10^{18}$ G, as expected in the crust region.  The
   authors of \cite{Lima2013} did not consider the possibility that at
   densities above $\rho_1$ new non-homogeneous regions would exist, as
   calculated in  \cite{Fang2017a}, using a dynamical spinodal
   approach.  For the region bounded by $\rho_1$ and $\rho_2$ in
   Fig. \ref{scheme1}, we will take results of Refs.~\cite{Fang2017,Fang2017a} to define the location of the
   non-homogeneous regions, since presently no other results are
   available that identify these regions.

We  consider two models which only differ in the
isovector properties: NL3$\omega\rho$ with the symmetry energy slope
$L=55$ and 88 MeV at saturation \cite{Pais2016vlasov}.   These values lie at
the average and top limit obtained  in \cite{Oertel2017} for the
symmetry energy slope at saturation from
constraints for nuclear properties and neutron star observations, $L= 58.7\pm 28.1$MeV.

\begin{figure}[!hbtp]
	\centering
	\includegraphics[width=0.5\textwidth]{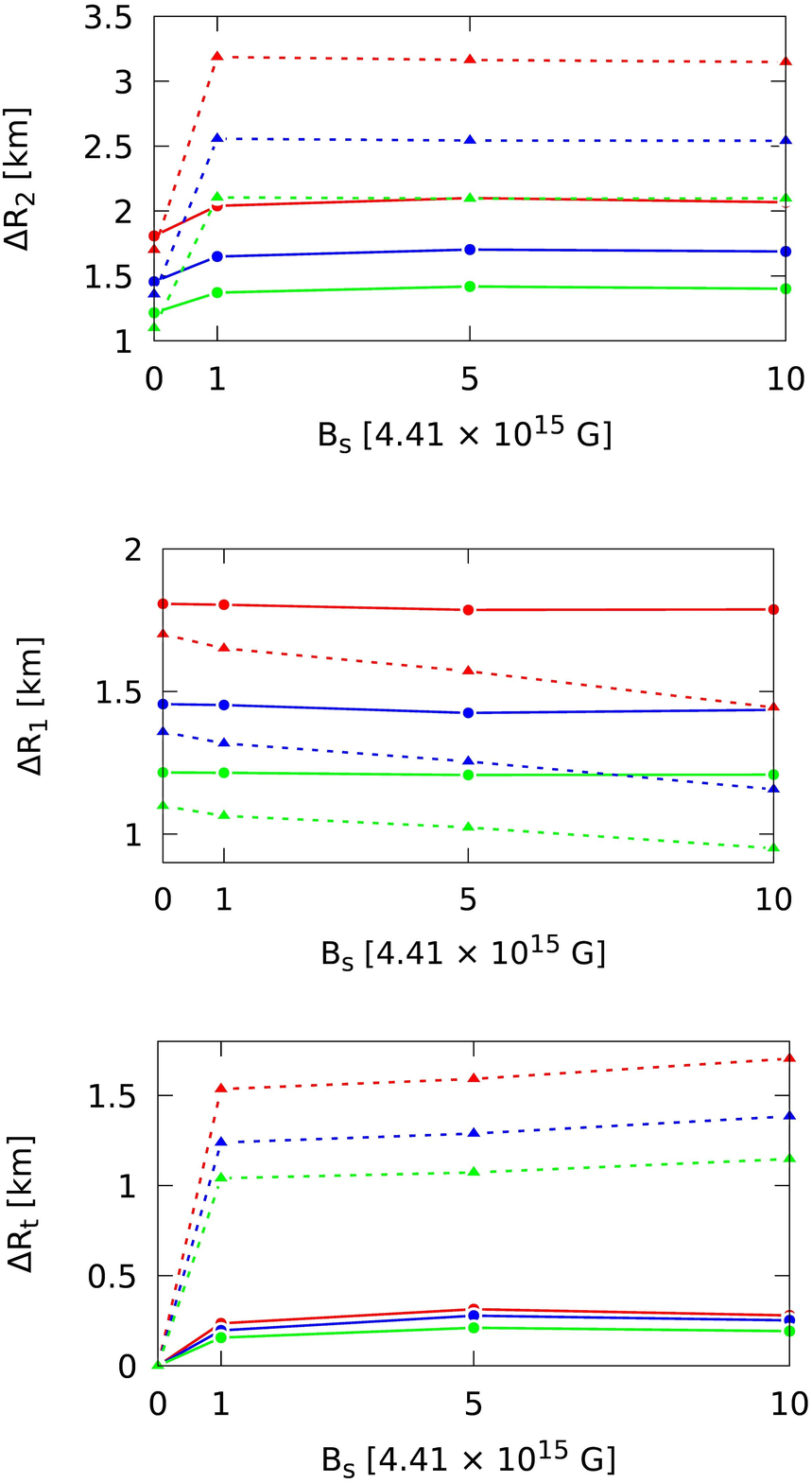}
	\caption{\label{mag1}(Color online) Effect of the magnetic field on the total size of the crust $\Delta R_2$ (top), on the crust without its extension, $\Delta R_1$, (middle), and on the extended region, $\Delta R_t$ (bottom). Full lines correspond to the model with $L=55$ MeV, whilst dashed lines are for the $L=88$ MeV model. The colours red, blue and green correspond to baryon masses $1.2 \mathrm{M}_\odot$,$1.5 \mathrm{M}_\odot$ and $1.8 \mathrm{M}_\odot$, respectively.}
\end{figure}

In order to study the effects of the magnetic field on the star crust, we
first analyze how the three quantities, $\Delta R_1$, $\Delta R_2$
and $\Delta R_t$, vary with the
radial component of the magnetic field measured at the surface (poles),  $B_s$. These
results are presented in Figure \ref{mag1} for stars with baryon masses 1.2,
1.4 and 1.8 $M_\odot$. On the top panel, we show
how the size of the crust is affected by the presence of the magnetic
field. We note that for $B_s=0$, the results for the two models
considered do not differ much from each other in comparison with the
case where $B_s \neq 0$. However, a difference does exist, as
discussed in \cite{Vidana2009}, where it was shown that the larger the
slope $L$, the smaller the  transition density to the
core. This, in turn, may reflect itself on the thickness of the crust:
in \cite{Grill2014}, it was found that a 
thinner crust corresponds to a larger $L$, when comparing NL3
($L=118$  MeV) with
NL3$\omega\rho$ with $L=55$ MeV. 

On the other hand, a much greater difference is verified for $B_s \neq
0$. This is because the value $\rho_2$, which defines the crust size,
depends on the proton fraction value considered. It was shown 
in \cite{Ducoin2008}  that the proton fraction at the crust-core
transition is determined by the slope of symmetry
energy, the smaller the $L$ the larger the proton fraction. A similar
conclusion was drawn in \cite{Grill2012} for the average proton
fraction at the  inner crust.  Therefore, even though both models
predict the same properties for symmetric nuclear matter, they will respond differently
with the inclusion of the magnetic field due to their  different symmetry energy properties. As a result, the model with $L=88$ MeV shows a much bigger
sensitivity to the increase of the magnetic field. The reason lies in the
fact that, for densities below saturation density, the fraction of
protons is smaller for larger values of $L$, and, therefore, more
sensitive to a given value of the magnetic field. It is also clear that
the smaller the star mass, the larger the effect of the magnetic field.

\begin{figure}[!hbtp]
	\centering
	\includegraphics[width=0.5\textwidth]{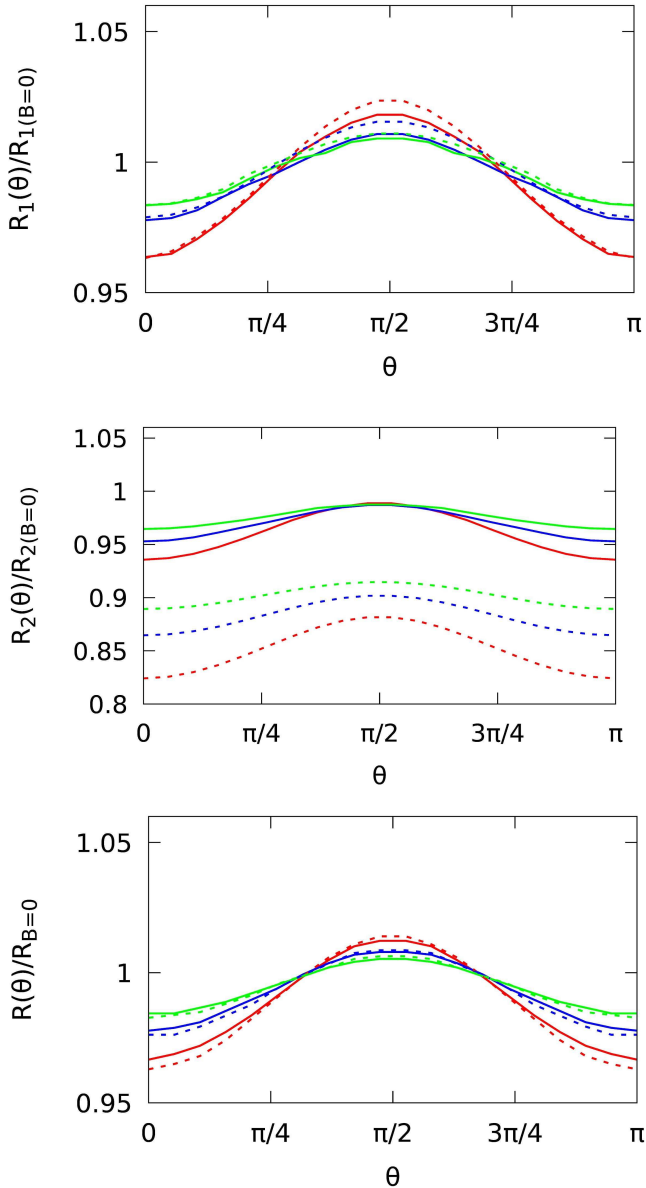}
	\caption{\label{mag2}(Color online) Normalised $R_1$, $R_2$ and $R$ as a function the polar angle, $\theta$.  Full lines correspond to the model with $L=55$ MeV, whilst dashed lines are for  the $L=88$ MeV model. The colours red, blue and green correspond to baryon masses $1.2 \mathrm{M}_\odot$,$1.5 \mathrm{M}_\odot$ and $1.8 \mathrm{M}_\odot$, respectively.  In each panel, the results obtained with $B_s=4.4 \times 10^{16}$ G are divided by the corresponding value at $B_s=0$ (notice that $R_2(B=0) = R_1(B=0)$). See text for more details.}
\end{figure}

\begin{figure}[!hbtp]
	\centering
	\includegraphics[width=0.5\textwidth]{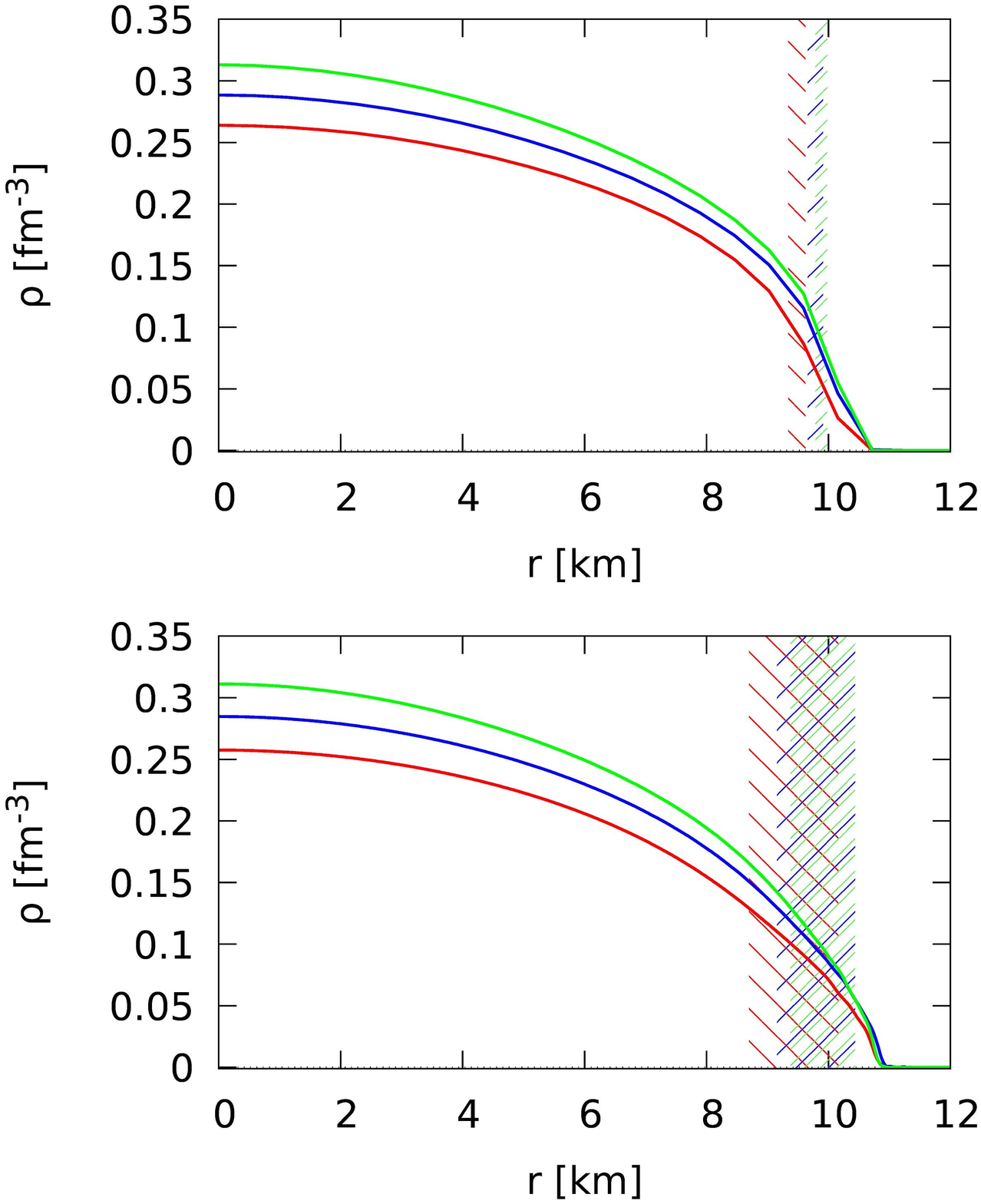}
	\caption{ \label{mag3}(Color online) Baryon density as function of the
          radial coordinate. The top panel corresponds to $L=55$ MeV
          whilst the bottom one corresponds to $L=88$ MeV.  The
          colours red, blue and green correspond to baryon masses $1.2
          \mathrm{M}_\odot$,$1.5 \mathrm{M}_\odot$ and $1.8
          \mathrm{M}_\odot$, respectively.  The vertical bands
          correspond to the transition zone at the crust-core
          transition.}  
\end{figure}

In the middle panel of Figure \ref{mag1}, it is shown how the size of
the crust without the extended zone varies with the magnetic
field. Here, the overall trend is a reduction on the size of the region,
as the magnetic field increases. Again we note that the model with the
larger $L$ is more affected by the increase of the magnetic field, a
conclusion that can also be reached by looking at the bottom panel of
the same figure, where we present the behavior of the extended crust
alone. It is important to note that this behavior is not monotonic,
which is a consequence of the discrete feature of the Landau levels
introduced by the magnetic field \cite{Fang2017}.

Concerning Fig.~\ref{mag1}, some comments are in
order: a) there is a large increase of the crust size when the
magnetic field increases from 0 to 4.4$\times 10^{15}$  G, but the
size of the crust is practically the same for 4.4$\times 10^{15}<B<
4.4\times 10^{16}$  G; b) the effect of the magnetic field is much
stronger if the model has a large symmetry energy slope; c) stars with
smaller masses are more strongly affected.

\begin{figure}[!hbtp]
	\centering
	\includegraphics[width=0.5\textwidth]{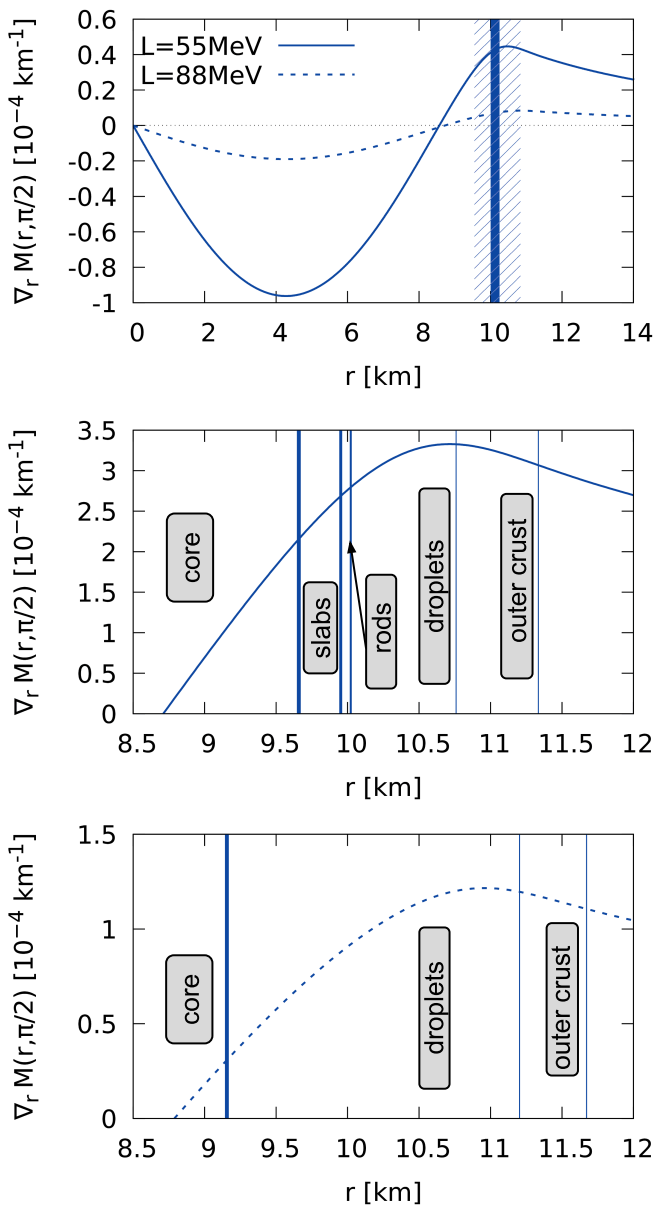}
	\caption{\label{mag4}(Color online) Gradient of the magnetic potential as function of the radial coordinate for a $M_b=1.5 \mathrm{M}_\odot$ star with $B_s= 4.4 \times 10^{16}$ G. Full lines correspond to the model with $L=55$ MeV, whilst dashed lines are for the $L=88$ MeV model. The two bottom panels show in more detail the region inside the crust, and the transition between the different pasta phases is signaled.}
\end{figure}

In Figure \ref{mag2}, we have used the results obtained with the
stronger magnetic field intensity ($B_s=4.41 \times 10^{16}$ G at the surface's pole) to
show how the width of  the crust varies along the polar angle
$\theta$. We have normalized the curves with the values obtained with
$B=0$: for both $R_1$ and $R_2$, we divided the values obtained with
$B_s=4.41 \times 10^{16}$ G by the corresponding values (i.e. same mass
and same $L$) obtained at $B = 0$. Since the values $\rho_1$ and
$\rho_2$ do not have any spatial dependency, the results that we
observe here are only consequence of the overall deformation of the
star induced by the magnetic field. In fact, the way the crust is
deformed is quite similar to the deformation of the radius (i.e.,
coordinate radius) of the star itself, as shown on the bottom panel of
the same figure, where the radius $R$ of the star is plotted versus the polar angle. Nonetheless,  it
becomes clear from Figure \ref{mag2} that the effect of the
magnetic field is much stronger in the $L=88$ MeV model: the
difference between the equatorial and polar radii is larger; and the
extended crust extends much more into the interior of the star. As
discussed before,  the magnetic field has a  stronger  effect on the
width of the crust of the less massive star: for the 1.2 $M_\odot$
star, the ratio between the equatorial radius is $\approx 5\%$ larger
then the polar radius, while for the 1.4 and 1.8 $M_\odot$ stars, this
difference is $\approx 2-3\%$. It is also interesting to notice that
the reduction of the radius at the pole is stronger than its increase
at the equator. This is also true for the thickness of the crust. The
middle panel shows that  the  location of the transition of the extended
crust-core is shifted towards the interior of the star for the model with
$L=88$  MeV, and  the star with the smallest mass. This shift is larger
at the equator, going up to more than 5\% (15\%) for the model with $55$
MeV (88 MeV). At the pole,  it  is not more than 1\% for the $L=55$ MeV  model, but rises to above 10\%  for the $L=88$ MeV  model.

This is also evident in Figure \ref{mag3}, where we plot the profile of each
star that we have considered and, in each of them, we identify the
extended zone, the region delimited by $R_2$ and $R_1$. By doing this,
we observe that the extended crust, which is itself a consequence of
the inclusion of the magnetic field,  is much  bigger for the
model with the larger $L$. 

We next analyze how the magnetic field potential varies inside the
star, and we discuss the localization of the points where its gradient,
proportional to the Lorentz force,  is
extreme and zero.

On the top panel of Fig. \ref{mag4}, we plot the radial component of
the gradient of the magnetic potential measured along the
plane $\theta = 0$ as a function of the radial coordinate. This quantity gives us the shape of the radial
component of the Lorentz
force inside the star, since $\boldsymbol{F_L} \sim \nabla M \left( r
  ,\theta \right)$.  At the equator,  the  gradient of the magnetic
potential function is zero at the neutral line of the poloidal
magnetic field \cite{Lander2011}. For polar angles close to the
equator,  the Lorentz force
verifies a sign change inside the star, as discussed in
\cite{Franzon2017}, because the lines of field are closed. 
It was shown in \cite{Lander2011}, where the authors have studied
instabilities in NS with poloidal magnetic fields, that  the most
unstable perturbations develop around the neutral line. 

We wanted to  ascertain whether the neutral line coincides with the extended
crust region, which should be taken into account when one considers
strong magnetic fields. For the models considered, we verified that
that does not occur, and, in fact, we obtained the  neutral line at
$r=R_n$, with $R_n/R\sim 0.8$, as predicted in \cite{Lander2011}. It has, however, been shown that stability  in a magnetized star is attained with both a poloidal and a
toroidal component, with the last one embedded inside the region
defined by the poloidal closed lines \cite{Haskell2008,Lander2009,Uryu2019}. We have included in Table \ref{Tab-neutral} the
position of the neutral line $R_n$, and  the extension of the crust
$R_2$ and $R_1$, as well as the NS radius $R$, for stars with masses
1.2, 1.4 and 1.8 $M_\odot$ described by models with $L=55$ and
$L=88$ MeV, and the surface magnetic field $B_s=4.41 \times 10^{15}$G. In the next section, we will discuss the effect of rotation on the neutral line.

\begin{table}[!t] 
	\centering
	\caption{The position of the inner crust boundaries
		$R_1$ and $R_2$, the NS radius $R$, and the neutral line $R_n$ measured along the equatorial plane, for different values of the surface magnetic field, $B_s$, and for stars with masses 1.2, 1.4 and 1.8 $M_\odot$, and described by models with $L=55$ and $L=88$ MeV. Note: $B^*=4.41 \times 10^{15}$G.}
	\begin{tabular}{cccccc}
		\hline\hline
		$L$ (MeV)  & M$_b$ $(\mathrm{M}_\odot)$ &$R_1$ (km)  & $R_2$ (km)  & $R$ (km) &  $R_n$ (km)\\ \hline
		\multicolumn{6}{c}{$B_s=B^*$}\\
		\hline
		\multirow{3}{*}{55}	& 1.2	& 9.987 &  9.752 &11.79 & 8.660	\\	
			& 1.5	& 10.14	& 9.944	&11.59&8.640	\\
			& 1.8	& 10.15	&	9.991&11.36 & 8.522 	\\ 
	    \hline
		 \multirow{3}{*}{88}	& 1.2	&  10.60	&9.067	& 12.25  &	8.946\\
			& 1.5	& 10.64	&	9.402&11.96 & 8.820 	\\ 
			& 1.8	& 	10.59& 9.545 &11.65 &8.632 	\\
		\hline
		\multicolumn{6}{c}{$B_s=5 \, B^*$}\\
		\hline
		\multirow{3}{*}{55}	& 1.2	& 10.06 & 9.729	&	  11.83 & 8.691	\\	
			& 1.5	& 10.17& 9.907	& 11.62& 8.658\\
			& 1.8	& 10.19	&9.955&	11.38	&8.536	\\
	    \hline
		\multirow{3}{*}{88}	& 1.2	& 10.72 &	9.131	&12.30	&8.982	\\
			& 1.5	& 10.74	&9.444 &	11.99	&8.842\\
			& 1.8	& 10.65	&9.573& 11.67	& 8.646	\\
			\hline
		\multicolumn{6}{c}{$B_s=10 \,B^*$}\\
		\hline
		\multirow{3}{*}{55}	& 1.2	&  10.17	&9.873	& 11.94  &	8.775\\
		& 1.5	& 10.25	& 10.01& 11.69	&8.711	\\
		& 1.8	& 	10.24&10.02	&11.43	&	8.576\\
		\hline
		\multirow{3}{*}{88}	& 1.2	&  10.98&	9.267	&12.44	&	9.084\\
		& 1.5	& 	10.91& 9.528&		12.07&	8.905\\
		& 1.8	& 	10.77& 9.625&		11.73& 8.688	\\
		\hline\hline
	\end{tabular}
	\label{Tab-neutral}
\end{table}

Besides the neutral line, the Lorentz force has two local extrema inside the NS, one located at
the core and the other one in the crust. We may assume that a maximum
of the Lorentz force inside the non-homogeneous region of the star may
cause more easily  matter to fracture or break. On the middle and
bottom panels of  Fig. \ref{mag4}, we identify the location of the pasta
phases in the crust region. In the case of the $L=88$ MeV model, no
inner crust configurations besides droplets exist. However, for the
$L=55$ MeV, the maximum of the Lorentz force occurs near the region of
rod-like configurations, which may be easier to deform.  The
localization of the transition between  pasta  phases in our work is only indicative, since
they have been obtained in a calculation that considered  the
possible formation of only five
different configurations. In a calculation that allows the appearance
of any kind of geometry as in
\cite{Schneider2014,Schneider2016,Caplan2018}, the extreme of the
Lorentz force would most probably fall in a region of non-spherical pasta phases.

\begin{figure}[t]
	\centering
	\includegraphics[width=0.5\textwidth]{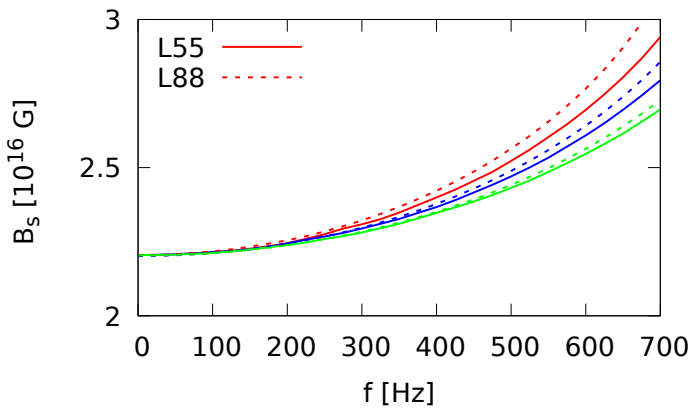}
	\caption{(Color online) Radial component of the magnetic field at the pole
          surface, $B_s$, as function of rotation frequency, $f$, for a fixed  magnetic dipole moment. Full lines correspond to the model with $L=55$ MeV, whilst dashed lines are for $L=88$ MeV. The colors red, blue and green correspond to baryon masses $1.2 \mathrm{M}_\odot$,$1.5 \mathrm{M}_\odot$ and $1.8 \mathrm{M}_\odot$, respectively.}
	\label{magrot1}
\end{figure}

\subsection{ Magnetised and rotating neutron stars}
\label{rotating}

The effects of rotation on the geometry of neutron stars are already
well documented \cite{Bocquet:1995je}, the major result being the flatness
of the star on the polar regions, an effect similar to that of the
polar magnetic fields discussed in the previous section.  In
\cite{Lander2011}, the authors showed that rotation
stabilizes the instabilities developed  in neutron stars with  a  poloidal  magnetic field
due to  perturbations.  Here we analyze how the profile of the Lorentz force in the equatorial plane is affected,
relatively to the crust,  when we take into account the effects of
rotation. In particular, we will determine the frequency above which
the neutral line does not exist.

It is important to notice that even though we fix the magnitude of the
magnetic field on the star by choosing the current function amplitude
($CFA$), which is equivalent to fixing the magnetic dipole moment, the magnitude
of the magnetic field measured at the pole surface $B_s$ is going to vary as we
increase the frequency.  This stems from the
fact that the angular velocity of the fluid and the magnetic function
are related by the fluid's conservation equation (\ref{EQMFINAL}). The
behavior of $B_s$ with the rotation frequency is shown in Fig. \ref{magrot1}, where the
radial component  of the magnetic field at the pole surface is plotted for  a fixed CFA value  (the one that gives, for each star, a field magnitude of $2.2 \times 10^{16}$ G when there is no rotation), using the two models of  the present study, and considering stars with masses 1.2, 1.4 and 1.8 $M_\odot$. We
conclude that larger magnetic field intensities are attained for the smaller mass
stars, and with a larger symmetry energy  slope. This happens because the proton fraction is bigger for the model with the larger $L$.

\begin{figure}[!t]
	\centering
	\includegraphics[width=0.5\textwidth]{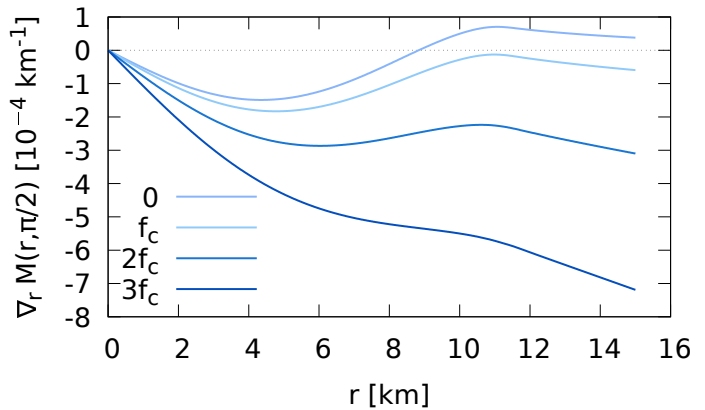}
	\caption{\label{mag5}(Color online) Gradient of the magnetic potential at
          the equator as a function of the radial coordinate for different values of the rotation frequency, and for the $L=55$ MeV model, and a $M_b=1.5 \mathrm{M}_\odot$ star with $B_s \approx 2.2 \times 10^{16}$ G.}
\end{figure}

As already discussed in \cite{Franzon:2016urz} and shown in the last
section, the magnetic potential function, $M$, may present a concave
shape and, thus, a local minimum. Since the Lorentz force is
proportional to the gradient of $M$, this minimum, at the equator,
corresponds to a line  of
points where the Lorentz force changes sign, and defines the neutral line. This means that there is a
region in which the magnetic field acts towards the center of the
star, and another one where the Lorentz force pushes outwards. A
change of the direction of the Lorentz force, if occurring in a fragile
region as the crust, could be  associated to the breaking of the
stellar crust and leading to flares. As discussed in the previous
section, in \cite{Lander2011} it was argued that in
the neighborhood of the neutral line large instabilities could develop
in a  star with a poloidal field.

Taking  into account the effects
of rotation, for each model and mass, there is a frequency (hereafter referred
to as \textit{critical frequency}, and designated by $f_{crit}$) at which the Lorentz force sign changes. This effect is shown in Figure \ref{mag5}, where we
present the radial component of $\nabla M$, measured along the
equatorial plane with
$\theta = \frac{\pi}{2}$. It is seen that for a frequency $f \gtrsim f_{crit}$, the
Lorentz force is always pointing outwards. The larger the frequency, the
stronger the Lorentz force.

In Table \ref{tab_freq1} we present the values of the critical frequency, $f_{crit}$, for the two models considered,  and for stars with $M_b=1.2 \mathrm{M}_\odot$, $M_b=1.5 \mathrm{M}_\odot$ and $M_b=1.8 \mathrm{M}_\odot$. We note that this so called critical frequency does not depend a lot on the model considered, but only on the baryonic mass of the star, and on the magnitude of the magnetic field. 
The critical frequencies obtained are all above 90 Hz. As shown, for
instance, in Ref.~\cite{Ho2013}, pulsars with strong magnetic fields have
periods of the order of 1 or larger. This means  that the poloidal
field inside these pulsars will always have a neutral magnetic line
and closed lines.

\begin{table}[!t] 
	\centering
	\caption{Frequency, $f_{crit}$, at which the neutral line disappears for the two models considered, and stars with different baryonic masses. The surface magnetic field is set to $B_s \approx 2.2 \times 10^{16}$ G.}
	\begin{tabular}{ccc}
		\hline\hline
		$L$ (MeV)  & M$_b$ $(\mathrm{M}_\odot)$ & $f_{crit}$ (Hz)\\ \hline
		\multirow{3}{*}{55}& 1.2	& 127\\
		& 1.5	& 109\\
		& 1.8	& 96\\
		\hline
		\multirow{3}{*}{88}	& 1.2	& 125\\
		& 1.5	& 108 \\
		& 1.8	& 96\\
		\hline\hline
	\end{tabular}
	\label{tab_freq1}
\end{table}

In Table \ref{tab_freq2} we show how the neutral line position is altered by the inclusion of  rotation for the models and masses previously considered. Similarly to what happens to the full coordinate radius of the star, the distance of the neutral line to the star centre increases with the frequency. The lower mass stars are the ones where this effect is more evident. On the other hand, for $f=50$Hz  the increase is roughly the same for the two models: $\sim 2.2 \%$ for the lower mass stars.

\begin{table}[!t] 
	\centering
	\caption{ The position of the inner crust boundaries $R_1$ and $R_2$, the NS radius $R$, and the neutral line $R_n$ measured along the equatorial plane, for different rotation frequencies, and stars with masses 1.2, 1.4 and 1.8 $M_\odot$, and described by models with $L = 55$ and $L = 88$ MeV. The surface magnetic field at the pole is $ B_s\approx 2.2 \times 10^{16}$ G.	}
	\npdecimalsign{.}
	\nprounddigits{5}
	\begin{tabular}{cccccc}
		\hline\hline
		$L$ (MeV)  & M$_b$ $(\mathrm{M}_\odot)$ &$R_1$ (km)  & $R_2$ (km)  & $R$ (km) &  $R_n$ (km)\\ \hline
		\multicolumn{6}{c}{$f=0$ Hz}\\
		\hline
		\multirow{3}{*}{55}	& 1.2	& 10.06 & 9.729 & 11.83 & 8.691	\\	
		& 1.5	& 10.17 & 9.907 & 11.62 & 8.658	\\
		& 1.8  & 10.19 & 9.955 & 11.38 & 8.536 	\\ 
		\hline
		\multirow{3}{*}{88}	& 1.2	& 10.72 & 9.131 & 12.30 & 8.981\\
		& 1.5	& 10.74 & 9.444 & 11.99 & 8.842 	\\ 
		& 1.8	& 10.65 & 9.573 & 11.67 & 8.646 	\\
		\hline
		\multicolumn{6}{c}{$f=0.1$ Hz}\\
		\hline
		\multirow{3}{*}{55}	& 1.2	& 10.06 & 9.729 & 11.83 & 8.691	\\	
		& 1.5	& 10.17& 9.907 & 11.62 & 8.658\\
		& 1.8  & 10.16 & 9.955 & 11.38 & 8.536	\\
		\hline
		\multirow{3}{*}{88}&1.2 	& 10.72 & 9.131 & 12.30 & 8.982	\\
		& 1.5	& 10.73 & 9.443 & 11.99 & 8.842\\
		& 1.8	& 10.65& 9.573& 11.67 & 8.646	\\
		\hline
		\multicolumn{6}{c}{$f=10$ Hz}\\
		\hline
		\multirow{3}{*}{55}	& 1.2	& 10.06 & 9.729 & 11.83 & 8.699	\\	
		& 1.5	& 10.17 & 9.907& 11.62 & 8.668 \\
		& 1.8	& 10.19 & 9.955 & 11.38 & 8.549	\\
		\hline
		\multirow{3}{*}{88}	& 1.2	& 10.73 & 9.131 & 12.30 & 8.990 	\\
		& 1.5	& 10.74 & 9.444 & 11.99 & 8.852\\
		& 1.8	& 10.65 & 9.573 & 11.67 & 8.659	\\
		\hline
		\multicolumn{6}{c}{$f=50$ Hz}\\
		\hline
		\multirow{3}{*}{55}	& 1.2	& 10.05 & 9.728 & 11.84 & 8.876\\
		& 1.5	& 10.18 & 9.914 & 11.63 & 8.917	\\
		& 1.8	& 10.19 & 9.960 & 11.39 & 8.872\\
		\hline
		\multirow{3}{*}{88}	& 1.2	& 10.73 & 9.133 & 12.31 & 9.183\\
		& 1.5	& 10.74 & 9.446 & 11.995 & 9.113\\
		& 1.8	& 10.65 & 9.576 & 11.67 & 8.991	\\
		\hline\hline
	\end{tabular}
	\npnoround
	\label{tab_freq2}
\end{table}

\begin{figure}[!hbtp]
	\centering
	\includegraphics[width=0.5\textwidth]{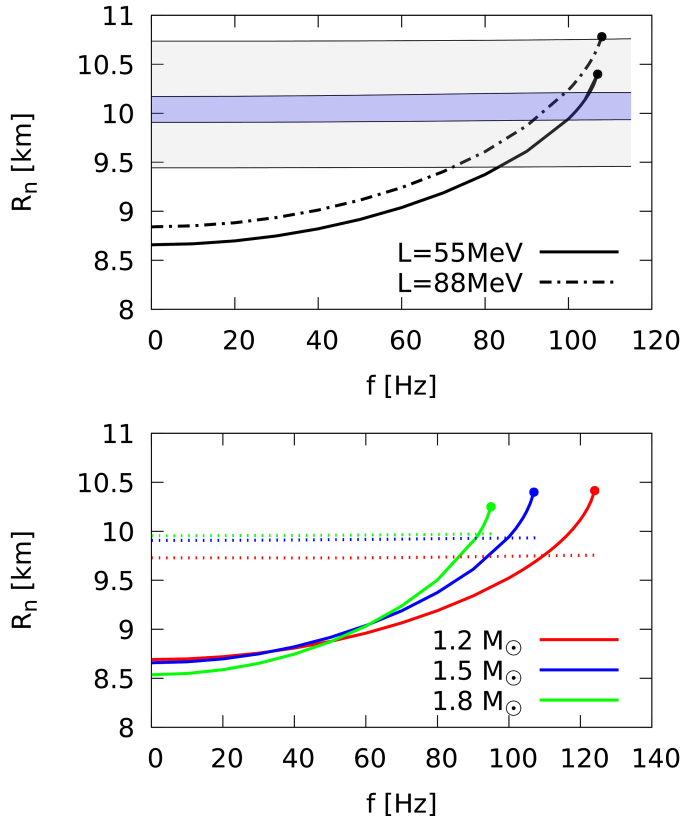}
	\caption{\label{fig_nl_rot}(Color online) Neutral line as function of the frequency. The top panel shows the results for a $1.5 M_{\odot}$ star, with $L=55$ MeV (solid line, blue horizontal band) and $L=88$ MeV (dashed line, grey horizontal band). The horizontal bands correspond to the extended crust. The bottom panel shows the results for the $L=55$ MeV model, for stars with $1.2$ (red), $1.5$ (blue) and $1.8$ (green) M$_{\odot}$. The horizontal lines correspond to the crust-core boundary, i.e. $R_2$. In all cases, the magnetic field at the surface is set to  $B_s \approx 2.2 \times 10^{16}$ G. }
\end{figure}

As already mentioned in the previous section, stars endowed with a poloidal magnetic field may show instabilities around the neutral line, and it is believed that rotation might cease those instabilities \cite{Lander2011}. Unlike the non-rotating case, Figure \ref{fig_nl_rot} shows that the neutral line can indeed fall inside the crust region, when the extended crust is taken into account. In the bottom panel of the same figure, we show, for the model with $L=55$ MeV, how the neutral line is affected by the frequency increase for different masses. We conclude that lower mass stars are much more sensitive to the effects of the frequency. The results are analogous for $L=88$ MeV, however, the neutral line enters the extended crust for smaller frequencies.  

\section{ Conclusions}
\label{conclusions}

 In this paper, we analyse how strong magnetic fields and rotation affect the inner crust of a NS. The inner crust is complemented with an extended crust which, as reported in \cite{Fang2016,Fang2017}, should be taken into consideration when strong magnetic fields are present. Part of our goal was to understand how models of the same family, but with different symmetry energy slope, $L$, compare when subject to extreme magnetic fields and rotation. Our results show that the larger the  slope of the symmetry energy $L$, the bigger  the sensitivity of the model regarding variations of the magnetic field, which is consistent with the fact that below saturation density, the fraction of protons is smaller for larger values of $L$, and above it is larger. This is particularly evident on the difference in the size of the extended crust.
The magnetic field may affect the different types of layers that exist in the crust. We verified that the Lorentz force has two local maxima, one of them localized in the region populated by pasta phases. This indicates that  the geometries more susceptible to break lie in a region where some of the strongest stresses occur.

Studies on the evolution of magnetic fields in neutron stars have reported the existence of a line inside the star, the neutral line, where the magnetic field is zero. These same studies indicate the existence of instabilities around this line \cite{Lander2011}, if a pure poloidal field is considered. If a mixed magnetic field configuration is assumed, the toroidal field lies on top of the poloidal neutral line \cite{Uryu2019}. We wanted to ascertain whether this line falls inside the inner crust, when one takes into account the extended crust. This was not verified for non-rotating stars, but the situation changes when one includes rotation. 
Given the richness of phenomena that occur at the region of the neutral line, it is expected that they will depend on the properties of matter that is present in this region.
It would be interesting to understand the role that pasta phases might have in connection to known astrophysical phenomena associated with magnetars.

\section*{ACKNOWLEDGMENTS}
This work was partly supported by the FCT (Portugal) Projects No. UID/FIS/04564/2019, UID/FIS/04564/2020 and POCI-01-0145-FEDER-029912, and by PHAROS COST Action CA16214. H.P. acknowledges the grant CEECIND/03092/2017 (FCT, Portugal). 

\bibliographystyle{apsrev4-1}

\end{document}